\documentclass[10pt]{article}
\usepackage[dvips]{graphicx}
\title{
{\bf Semiclassical approaches to 
controlling chemical reaction dynamics}} 
\author{Hiroshi FUJISAKI,
Yoshiaki TERANISHI$^{\spadesuit}$,
Alexey KONDORSKIY, 
\\
and Hiroki NAKAMURA\footnote{e-mail: fujisaki@ims.ac.jp, 
tera@apr.jaeri.go.jp, kondor@ims.ac.jp, nakamura@ims.ac.jp} 
\\
{\it Department of Theoretical Studies,
Institute for Molecular Science}
\\
{\it $^{\spadesuit}$Japan Atomic Energy Research Institute}}
\date{}
\begin{document}
\maketitle

\begin{abstract}

We propose to use semiclassical methods 
to treat laser control problems of chemical reaction dynamics.
Our basic strategy is as follows: 
Laser-driven chemical reactions are considered 
to consist of two processes. One is the wavepacket 
propagation on an adiabatic potential energy surface (PES), 
and the other is the electronic transition between PES's. 
Because the latter process is mathematically equivalent 
to nonadiabatic transitions between Floquet (dressed) states, 
we can control such a process using the semiclassical Zhu-Nakamura 
theory for nonadiabatic transitions. 
For the former process, 
we incorporate semiclassical propagation methods 
such as the Herman-Kluk propagator into  
optimization procedures like optimal control theory. 
We show some numerical examples for our strategies.
%discuss the difficulty to implement the semiclassical 
%propagator into optimal control theory.
We also develop a semiclassical direct algorithm to treat 
the adiabatic propagation and nonadiabatic transitions as a whole.  

\end{abstract}

\section{Introduction}

Controlling molecular dynamics is one of the intriguing 
targets in chemical physics.
Since the works of Brumer and Shapiro, Tannor and Rice, and others, 
laser control of molecular dynamics has been developed 
relatively well for simple molecular processes \cite{RZ00}.
Theoretically, 
in order to obtain an optimal field producing a desired product, 
optimal control theory (OCT) is one of the most 
natural and flexible vehicles \cite{RZ00}. It assumes a certain functional 
which should be maximized, from which one can derive 
equations of motion which in turn determines the optimal field.
The equations including Schr\"odinger equation must be 
solved iteratively in general, 
hence its numerical cost becomes huge for large molecules.
Although there exist some calculations for reduced dimensionality 
models of large molecules, 
it can be very difficult to control even a triatomic molecule 
quantum mechanically within the present state of art. 
%(algorithms and computers).    
%On the other hand, the optimal fields provided by OCT 
%are often too complicated to be realized in experiment,
%and too hard to be understood from theoretical point of view.
Some strategies 
to circumvent this difficulty are strongly desired.

Our basic strategy is as follows: 
Laser-driven chemical reactions are considered 
to consist of two processes. One is the wavepacket propagation 
on an adiabatic potential energy surface (PES), 
and the other is the electronic transition between PES's.
Because the latter process is mathematically 
equivalent to nonadiabatic 
transitions between Floquet (dressed) states, 
we can control such a process using the semiclassical Zhu-Nakamura 
theory for nonadiabatic transitions \cite{ZTN01}.
(It has been shown in Refs.~\cite{ZTN01,NTN00,NTN02} 
that this strategy is very successful.)
In Sec.~\ref{sec:HI}, we shall show another example of our strategy.
A new thing here is that we treat a real diatomic molecule, HI,  
with its real PES information. We shall use the idea of 
{\it complete reflection} to control HI photodissociation to 
two different channels. A complication arises due to 
existence of three electronically excited states \cite{FTN02}. 

For controlling the wavepacket dynamics on a PES, 
on the other hand, 
%there is no such simple strategy, and 
we need to employ 
a general optimization procedure like OCT.
There have been many applications of OCT in the literature; 
however, such works are often limited to rather small systems 
with less than three degrees of freedom.
This is because the conventional methods of wavepacket propagation 
using {\it grids} are limited to rather small systems.
{\it Semiclassical wavepacket propagation methods}
are believed to overcome this kind of difficulty with use of 
classical trajectories.
Hence it is very natural to examine the possibility 
of combining the semiclassical wavepacket propagation methods with OCT.
%This is discussed in Sections~\ref{sec:OCT} and \ref{sec:semi}.
In Sec.~\ref{sec:OCT}, 
we employ one of semiclassical wavepacket propagation methods,
the initial value representation (IVR) method using the Herman-Kluk
propagator \cite{MB00}, and incorporate 
it into OCT. It is known that the IVR method 
can approximately reproduce the full quantum results 
for many molecular systems, and its computational scalability 
is much less than that of the corresponding 
quantum calculations for large molecules. 
In addition, we can understand the ``physical aspects'' of the
optimal field because the IVR method represents
the result as a sum of the
quantities calculated from classical trajectories. 
%Thus the OCT combined with the IVR method can be a candidate to resolve the 
%above difficulties of usual OCT.
%Some preliminary results using one-dimensional systems 
%will be presented. 

The effectiveness of the IVR method has been mainly tested for single 
PES systems, i.e., systems without nonadiabatic transitions.
(For exceptions, see Ref.\ \cite{SM97}.)
On the other hand, we have the semiclassical Zhu-Nakamura theory 
of nonadiabatic transition \cite{ZTN01}, 
and it is shown that 
it can be used to control even wavepacket excitation 
between PES's under some conditions \cite{NTN02}.
In Sec.~\ref{sec:semi}, we examine 
the possibility to combine this kind of semiclassical theory 
of nonadiabatic transitions with the 
semiclassical wavepacket propagation method. 
%(which might widen the applicability 
%of the above semiclassical implementation of OCT for single potential 
%surface systems) 

\section{Controlling a molecule by complete reflection: 
Application to HI photodissociation}
\label{sec:HI}

We shall apply the strategy using complete reflection \cite{NTN00}
to HI photodissociation. 
The {\it ab initio} potential energy curve (PEC) information is reproduced in 
Fig.~2 of \cite{FTN02}. There are three electronically 
excited states, and their transition dipole moments depend on the 
internuclear distance significantly.

When applying a stationary laser field,
we can consider this system as a curve-crossing 
system: one curve from the electronically ground state 
crosses with the other three curves from the 
electronically excited states. 
At crossings nonadiabatic transitions occur and such 
a phenomenon can be accurately treated by the 
semiclassical ZN formula.
If the energy of a vibrational state satisfies 
a resonant-like condition, % on an upper adiabatic PEC, 
then the phenomenon, complete reflection, occurs:
such a wavepacket cannot escape from the upper adiabatic PEC
to the lower adiabatic PEC. (The lifetime is infinite.)
This strategy 
%one vibrational excitation and one electronic excitation,
can be used to control photodissociation and 
some model calculations were done in \cite{NTN00}.

%\begin{figure}[htbp]
%\hfill
%\begin{center}
%%\hspace{1.2cm}
%\begin{minipage}[b]{.45\linewidth}
%\includegraphics[width=0.8\linewidth]{fig/pes.eps}
%\end{minipage}
%%\hspace{6mm}
%\begin{minipage}[b]{.45\linewidth}
%\includegraphics[width=0.8\linewidth]{fig/dipole.eps}
%\end{minipage}
%\caption{
%{\sf 
%(a): Ab initio PECs for HI.
%Unit for the vertical axis is cm$^{-1}$.
%(b): Ab initio transition dipole moments between the electronically 
%ground and excited states. Taken from \cite{Balakrishnan}.
%Unit for the vertical axis is au.
%}
%}
%\label{fig:pes}
%\end{center}
%\end{figure}

\begin{figure}[htbp]
\hfill
\begin{center}
%\hspace{1.2cm}
\begin{minipage}[b]{.45\linewidth}
\includegraphics[width=1.0\linewidth]{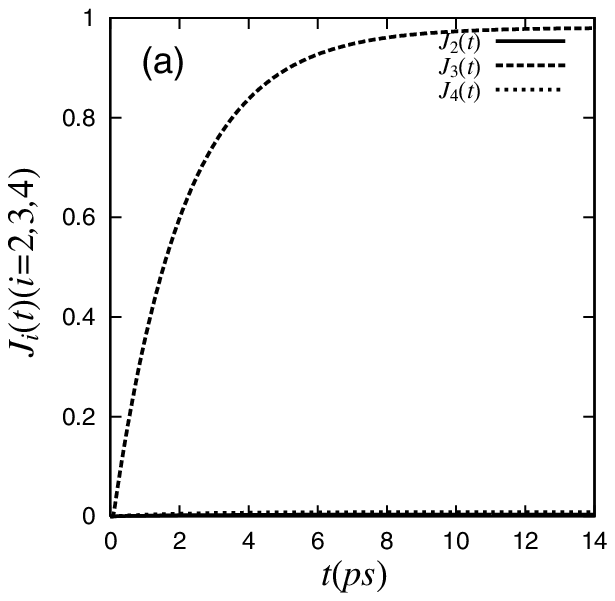}
\end{minipage}
\hspace{6mm}
\begin{minipage}[b]{.45\linewidth}
\includegraphics[width=1.0\linewidth]{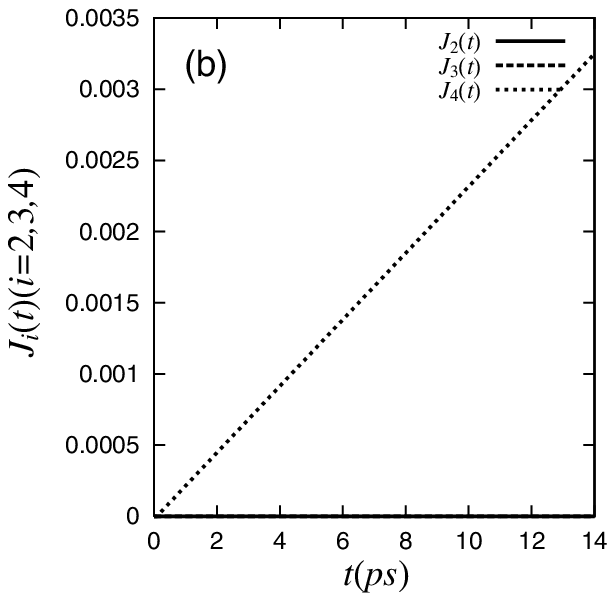}
\end{minipage}
\caption{
{\sf 
Time variation of the time-integrated fluxes $J_i(t)$ ($i=2,3,4$ corresponds
to three electronically excited states) for 
the initial vibrational state with (a) $v=4$ 
and (b) $v=0$. 
The laser frequency is $\hbar \omega \simeq 3.58$ eV.
}
}
\label{fig:flux_time}
\end{center}
\end{figure}

To control HI photodissociation, we need to control 
three crossings {\it at the same time}. Of course, this 
is {\it mathematically} impossible because the complete reflection
occurs at a certain energy point, but it might work 
{\it in practice} at some time interval.
As shown in Fig.~\ref{fig:flux_time}(a), 
at a certain energy ($\hbar \omega \simeq 3.58$ eV) and up to 
a certain time interval (14 ps), 
the two fluxes ($J_2(t), J_4(t)$) 
corresponding to two electronically excited PEC's 
connected to H+I channel nearly equal zero 
because this energy satisfies the complete reflection 
condition for {\it both} states to a good extent.
In this case, we have to prepare the vibrationally excited state
with $v=4$ ($v=5$ is better from 
experimental points of view \cite{Kawasaki}, 
but the qualitative result is the same).  
In addition, the contribution from $v=0$ component, 
which might remain as a major component even 
after infrared excitation of the 
vibrational state, is very small as shown in 
Fig.~\ref{fig:flux_time}(b).
This means that the H+I$^*$ photodissociation can be 
nearly completely achieved by using the vibrational state with 
$v=4$ and the stationary laser field with $\hbar \omega \simeq 3.58$ eV.
(In contrast, it is easy to achieve H+I photodissociation because 
only single excited PEC is asymptotically connected to H+I$^*$ which 
should be blocked.)

We have not included the effect of rotation and temperature, 
which degrade the quality of control, 
but the actual experimental condition is expected to be 
much similar to our calculation \cite{Kawasaki}.

\section{Semiclassical implementation of optimal control theory}
\label{sec:OCT}

\begin{figure}[htbp]
\hfill
\begin{center}
%\hspace{1.2cm}
\begin{minipage}[b]{.43\linewidth}
\includegraphics[width=1.0\linewidth]{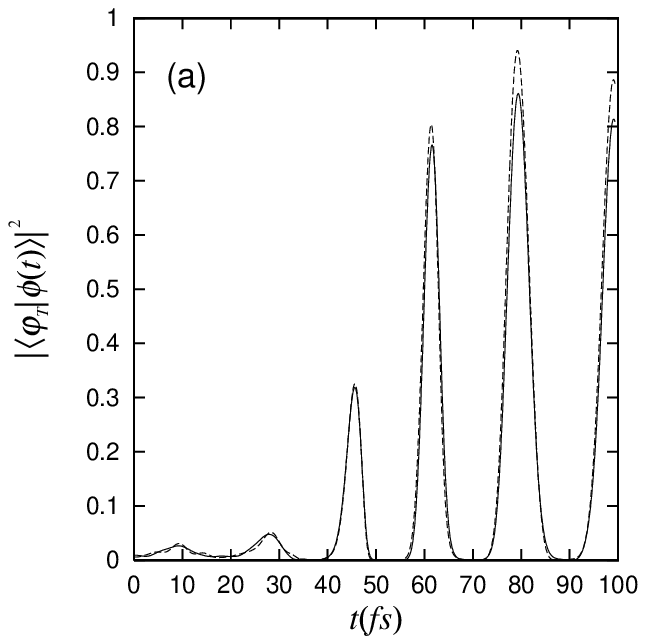}
\end{minipage}
\hspace{6mm}
\begin{minipage}[b]{.43\linewidth}
\includegraphics[width=1.0\linewidth]{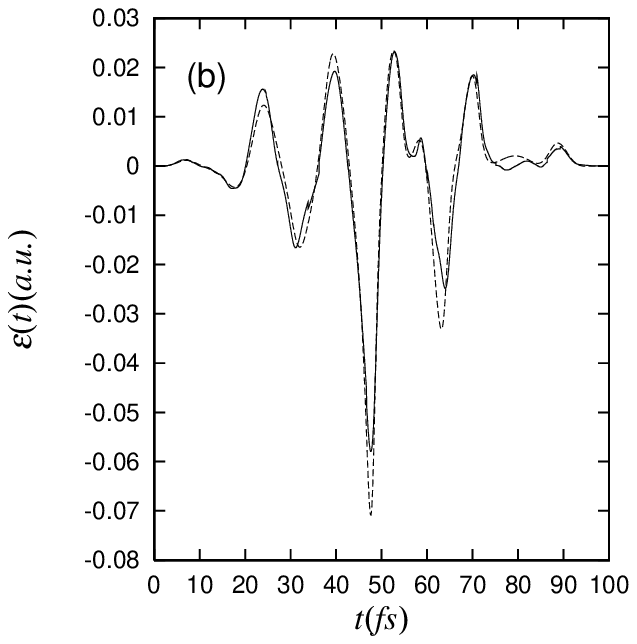}
\end{minipage}
\caption{
{\sf 
(a) Semiclassical (solid line) and quantum (dashed line) calculations 
of the overlap integral $|\langle \varphi_T| \phi(t) \rangle|^2$ where 
$|\varphi_T \rangle$ is the target state. 
(b) Semiclassical (solid line) and quantum (dashed line) optimal fields
obtained after one iteration.
}
}
\label{fig:OCT}
\end{center}
\end{figure}

In the previous section, we have shown that the electronic 
transition can be controlled with the help of the 
semiclassical ZN formula.
As mentioned in the introduction, however, 
laser-driven chemical reactions consist of 
two processes, and we need to control 
not only electronic transitions 
but also wavepacket propagation. 
There seems to be no simple semiclassical methods to deal with 
the latter, so we examine the possibility to combine 
the optimal control theory (OCT) with a semiclassical wavepacket 
propagation method with the Herman-Kluk propagator \cite{MB00}.
%\begin{eqnarray}
%G^{\rm HK}( \mbox{\boldmath $x$}, \mbox{\boldmath $x$}_0,t)
%&=&
%\left( \frac{1}{2 \pi \hbar} \right)^f
%\int d  \mbox{\boldmath $p$} \int d \mbox{\boldmath $q$}
%\, C_{pqt} \, e^{i S_{pqt}/\hbar}
%g_{p_t q_t}(\mbox{\boldmath $x$} ) 
%g^*_{pq}(\mbox{\boldmath $x$}_0),
%\\
%C_{pqt}
%&=&
%\pm
%%\left( \frac{1}{2 \pi \hbar} \right)^f 
%\left \{ {\rm det}
%\left [ \frac{1}{2} 
%\left ( 
%\frac{\partial \mbox{\boldmath $p$}_t}{\partial \mbox{\boldmath $p$}}
%+\frac{\partial \mbox{\boldmath $q$}_t}{\partial \mbox{\boldmath $q$}}
%-2 i \hbar \gamma \frac{\partial \mbox{\boldmath $q$}_t}
%{\partial \mbox{\boldmath $p$}}
%-\frac{1}{2 i \hbar \gamma} 
%\frac{\partial \mbox{\boldmath $p$}_t}{\partial \mbox{\boldmath $q$}}
%\right)
%\right]
%\right \}^{1/2},
%\\
%S_{pqt}
%&=&
%\int_0^t d\tau
%[\mbox{\boldmath $p$}_{\tau} 
%\cdot 
%\dot{\mbox{\boldmath $q$}}_{\tau} 
%-H(\mbox{\boldmath $p$}_{\tau},\mbox{\boldmath $q$}_{\tau})],
%\\
%g_{p q}(\mbox{\boldmath $x$})
%&=& \left( \frac{2 \gamma}{\pi} \right)^{1/4} 
%e^{-\gamma( \mbox{\boldmath $x$} -\mbox{\boldmath $q$})^2 
%+i \mbox{\boldmath $p$} \cdot 
%(\mbox{\boldmath $x$}  -\mbox{\boldmath $q$})/\hbar}
%\end{eqnarray}
%where $f$ is the number of degrees of freedom 
%of the system,
%$C_{pqt}$ represents instability of 
%the trajectories, 
%%(as is the case with 
%%the prefactor in the Van-Vleck-Gutzwiller propagator), 
%$S_{pqt}$ is the classical action which 
%connects $(q,p)$ and $(q_t,p_t)$,
%and $\gamma$ is a free (adjustable) parameter.
As an OCT procedure, we take the 
Zhu-Botina-Rabitz method \cite{ZBR98}, and 
substitute the semiclassical propagator 
into the control scheme.
The numerical result for HD$^+$ molecule is shown in 
Fig.~\ref{fig:OCT}: The control here is to shift the vibrational
ground state to 1 a.u. away from the equilibrium position, 
the target time $T$ is 100 fs,
and the so called penalty factor $\alpha$ is 0.01.

As can be seen, the semiclassical result quantitatively 
agrees with the quantum result.
Unfortunately, however, this straightforward semiclassical 
calculation needs more computational time compared with 
the corresponding quantum calculation!
This is because the optimal field expressed as 
\begin{equation}
%\nonumber
\varepsilon(t)=-\frac{1}{\alpha} {\rm Im} 
\left[ 
\langle \phi(t)|\chi(t) \rangle 
\langle \chi(t)| \mu |\phi(t) \rangle 
\right]
\end{equation}
requires ${\cal O}(N^2)$ calculations 
where $N$ is the number of the trajectories.
Here $|\phi(t) \rangle$ and $|\chi(t) \rangle$ are 
semiclassically calculated wavefunctions with 
computational cost ${\cal O}(N)$;
so
$\langle \phi(t)|\chi(t) \rangle$ and
$\langle \chi(t)| \mu |\phi(t) \rangle$ amount to 
such an order of calculation in general. 
($\mu$ is the transition dipole moment.)
One might expect that we can reduce the cost 
by decreasing the width of the Gaussian 
wavepacket in the HK propagator;
but it turns out that the overall 
efficiency cannot be improved.
The same difficulty appears when 
one semiclassically calculates general correlation 
functions as analyzed by Miller \cite{Miller01}.
%but when the control purpose 
%is to shift the ground state to the other place,
%then the $\gamma$ should be determined from the 
%potential shape, and it should not be changed.

We are now planning to use a sort of {\it genetic algorithm} \cite{JR92}
to circumvent this difficulty since the optimal fields in the genetic 
algorithm procedure are represented as ``genes'', 
and there is no such a scaling problem.
(The initially guessed field can be obtained from 
the concept of nonadiabatic transition between vibrational states.)

\section{Semiclassical IVR method for nonadiabatic transitions}
\label{sec:semi}

The semiclassical methods mentioned above {\it separately} treat 
the processes of adiabatic propagation on a single adiabatic PES 
%(using the Herman-Kluk method) 
and those of electronic transition.
 %(using ZN formula) separately. 
This means that the whole process should be divided into three steps:
(i) propagation on an initial PES, 
(ii) nonadiabatic transition, and
(iii) propagation on a final PES.
%The propagation (i) and (iii) are treated by the Herman-Kluk 
%method, and 
The optimal field should be designed so that 
no unnecessary interference occurs between the steps.
In order to overcome such a restriction and to use the 
unique pulse to control the process, 
semiclassical IVR methods for nonadiabatic transitions
should be constructed.
One of the purposes of this work is to develop such a method
which  generalizes the ordinary Herman-Kluk propagator 
for the case of several electronic states.
For other studies in this direction, see \cite{SM97}.

\begin{figure}[htbp]
\hfill
\begin{center}
%\hspace{1.2cm}
\begin{minipage}[b]{.45\linewidth}
\includegraphics[width=1.0\linewidth]{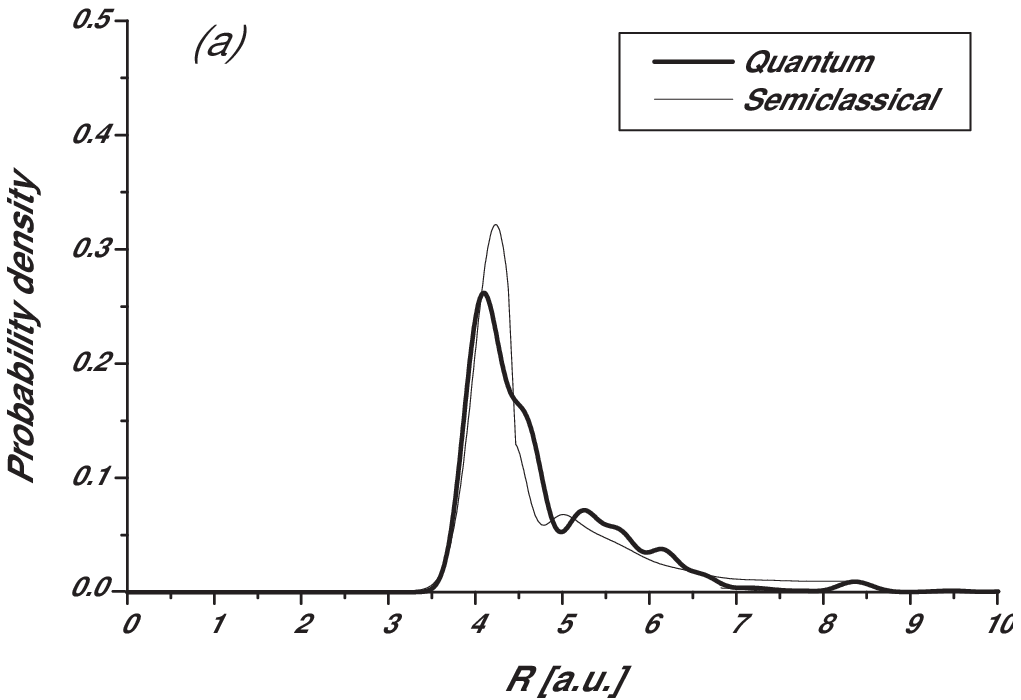}
\end{minipage}
\hspace{6mm}
\begin{minipage}[b]{.45\linewidth}
\includegraphics[width=1.0\linewidth]{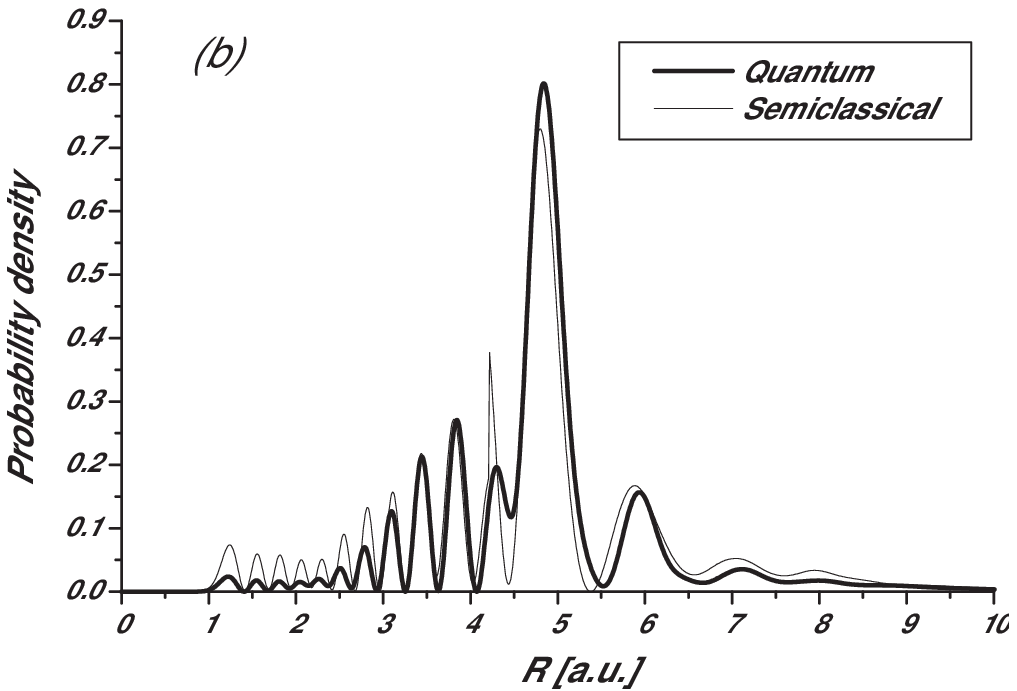}
\end{minipage}
\caption{
{\sf 
(a) Final wavepacket on the 2p$\sigma_u$ state
calculated quantum mechanically (thick line) 
and semiclassically (thin line).
The total population: 
Quantum =0.30; Semiclassical =0.28.
(b) The same on the 1s$\sigma_g$ state.
The total population: 
Quantum =0.70; Semiclassical =0.72.
}
}
\label{fig:semi}
\end{center}
\end{figure}

The research is still under way
and preliminary results for a 1D case are presented. 
The model system employed is H$_2^+$ molecule
composed of 1s$\sigma_g$ and 2p$\sigma_u$ states.
An initial wavefunction is assumed to be a Gaussian of width 0.5 a.u. 
with center at 5 a.u. on the 1s$\sigma_g$ state. 
The external field is assumed to be a linearly polarized femtosecond 
pulse with the Gaussian profile of FWHM = 10 fs. 
The total duration of wavepacket propagation is 20 fs. 
The wavelength of the femtosecond pulse is 515 nm and 
the crossing point of the dressed electronic states 
is located at 4 a.u. 
The results for the wavepacket propagation are shown
in Fig.~\ref{fig:semi}.

\section{Outlook}

%We shall consider the control problems of chemical 
%reactions from more general points of view.
We have advocated the use of semiclassical methods for 
control problems in view of 
its general capability to treat multidimensional 
systems with less computational cost. % {\it in principle}.
If the dimensionality of the system is 
high, the effects of ``chaos'' might also affect
the quality of control. We have shown that 
a chaotic system called a kicked rotor
can be controlled by using the Zhu-Botina-Rabitz 
method even when the system becomes strongly chaotic 
in the classical limit \cite{TFM03}.
It should be confirmed whether 
this is generic or not for multidimensional systems.
For much larger systems, even semiclassical calculations 
themselves are formidable, and in such a situation we 
need to use some kind of mixed quantum-classical 
methods. They are very practical ways to treat multidimensional 
phenomena like vibrational relaxation in solvents \cite{Okazaki},
so it is interesting to control such phenomena with use of 
mixed quantum-classical methods. 
%\\
%\\
%One of the authors (H.F.) thanks K.\ S.\ Ikeda and G.\ V.\ Mil'nikov 
%for critical comments.

\end{document}